\documentclass[a4paper]{article}
\usepackage{amsmath,amsthm,amssymb}
\numberwithin{equation}{section}

\newtheorem{definition}{Definition}[section]
\newtheorem{remark}{Remark}[section]
\newtheorem{lemma}{Lemma}[section]
\newtheorem{theorem}{Theorem}[section]
\newtheorem{corollary}{Corollary}[section]

\newcommand{\bbC}{\mathbb{C}} 

\newcommand{\g}{\ensuremath{\mathfrak{g}}}

\renewcommand{\sl}{\ensuremath{\mathfrak{sl}}}

\newcommand{\osp}{\ensuremath{\mathfrak{osp}}}

\renewcommand{\b}{\beta}
\newcommand{\betheop}{\hat{\beta}}

\renewcommand{\k}{\kappa}
\renewcommand{\l}{\lambda}
\newcommand{\bl}{\boldsymbol{\lambda}} 
\renewcommand{\L}{\Lambda}
\newcommand{\bmu}{\boldsymbol{\mu}} 
\renewcommand{\o}{\omega}
\renewcommand{\O}{\Omega}

\newcommand{\te}{\theta}

\newcommand{\pop}{\hat{p}}

\newcommand{\tr}{\operatorname{Tr}}
\newcommand{\cH}{\ensuremath{\mathcal{H}}}

\newcommand{\Lxi}{\mathcal{L}_{\xi}}
\newcommand{\x}{\times}
\newcommand{\ox}{\otimes}
\newcommand{\LI}{\ensuremath{\underset{1}{L}}}
\newcommand{\LII}{\ensuremath{\underset{2}{L}}}
\newcommand{\xp}{X^{+}}
\newcommand{\xm}{X^{-}}

\newcommand{\xmd}{(X^{-})^{\ast}}

\begin{document}

\begin{titlepage}

\begin{center}
{\bf\huge $\sl_{2}$ Gaudin model with jordanian twist}
\vskip 1.5 cm

{\sc N. ~Cirilo Ant\'onio}\\
\medskip
{\it Departamento de Matem\'atica, Instituto Superior T\'ecnico,\\
UTL, Lisbon, Portugal}\\
E-mail address: nantonio@math.ist.utl.pt\\
\vskip 1cm
{\sc N. ~Manojlovi\'c}\\
\medskip
{\it Departamento de Matem\'atica, F.C.T.,
Universidade do Algarve\\ Campus de Gambelas, 8005--139 Faro,
Portugal}\\
E-mail address: nmanoj@ualg.pt\\
\vskip 1.5cm

\end{center}

\begin{abstract}
$\sl_{2}$ Gaudin model with jordanian twist is studied. This system can be obtained as the semiclassical limit of the XXX spin chain deformed by the jordanian twist. The appropriate creation operators that yield the Bethe states of the Gaudin model and consequently its spectrum are defined. Their commutation relations with the generators of the corresponding loop algebra as well as with the generating function of integrals of motion are given. The inner products and norms of Bethe states and the relation to the solutions of the Knizhnik-Zamolodchikov equations are discussed.
\end{abstract}

\end{titlepage}

\clearpage \newpage

\section{Introduction}

The Quantum Inverse Scattering Method (QISM) was largely created by L. D. Faddeev and his school at St.~Petersburg as a quantum counterpart of the Classical Inverse Scattering Method \cite{faddeev1}\nocite{kulish10,sts}--\cite{Korepin1993uc}. Classifying quantum  solvable systems with respect to the underlying dynamical symmetry algebras, one could say that the Gaudin models \cite{gaudin2,gaudin} can be seen as the simplest ones being based on loop algebras. Their hamiltonians are related to classical r-matrices, 
\begin{equation}
\label{Gaudinham}
H^{(a)} = \sum_{b\neq a}^{N} r_{ab}(z_{a}-z_{b}).
\end{equation}
The condition of their commutativity $[H^{(a)},H^{(b)}]=0$ is nothing else but the classical Yang-Baxter equation 
\begin{equation}
\label{cYBE}
[ r_{ab}(z_{a}-z_{b}), r_{ac}(z_{a}-z_{c}) + r_{bc}(z_{b}-z_{c})] + [r_{ac}(z_{a}-z_{c}),r_{bc}(z_{b}-z_{c})]=0
\end{equation}
where $r$ is antisymmetric and belongs to the tensor product $\g\ox\g$ of a Lie algebra $\g$, or its
representations and the indices fix the corresponding factors in the $N$-fold tensor product of this
algebra. The Gaudin models based on the classical r-matrices of simple Lie algebras attracted a lot of attention \cite{sklyanin4}\nocite{sklyanin2,Babujian1994ba,frenkel}--\cite{Reshetikhin:1995sl}. Their spectrum and corresponding eigenfunctions were obtained using different methods such as coordinate and algebraic Bethe ansatz, separated variables, etc. The correlation functions of the $\sl_{2}$ Gaudin system were calculated by means of the Gauss factorization \cite{sklyanin2}. A connection between the Bethe vectors of the Gaudin models for simple Lie algebras to the solutions of the Knizhnik--Zamolodchikov equation was established in \cite{Babujian1994ba}\nocite{frenkel}--\cite{Reshetikhin:1995sl}. An explanation of this connection based on Wakimoto modules at critical level of the underlying affine algebra was given in \cite{frenkel}.

The algebraic Bethe Ansatz for the Gaudin model based on the $\sl_{2}$ invariant classical r-matrix deformed by the constant jordanian r-matrix was postulated in \cite{Kulishjs}. Following the ideas used in the case of the $\osp(1|2)$ trigonometric Gaudin model \cite{manoj}, Kulish noticed that the similarity transformation by $\exp(\alpha \xp)\ox\exp(\alpha \xp)$ on the $\sl_{2}$ trigonometrical classical r-matrix 
\begin{equation*}
\begin{split}
r_{trig}(\l) &=\frac{e^{-\l}}{\sinh(\l)}r^{(+)}+ \frac{e^{\l}}{\sinh(\l)}r^{(-)}\\
&=\frac{e^{-\l}}{\sinh(\l)}\left(\frac{1}{2}h\ox h + 2\xp\ox\xm\right)\\
&+\frac{e^{\l}}{\sinh(\l)}\left(\frac{1}{2}h\ox h + 2\xm\ox\xp\right),
\end{split}
\end{equation*}
setting $\l\to \epsilon\l$, $\alpha\to\frac{\xi}{2\epsilon}$ and after the scaling limit
$$
\lim_{\epsilon\to 0} \epsilon\, r_{trig}(\epsilon\l) =\frac{1}{\l}(h\ox h + 2(\xp\ox\xm+\xm\ox\xp)) + \xi (h\ox \xp - \xp\ox h),
$$
yields the $\sl_{2}$-invariant classical r-matrix deformed by the jordanian r-matrix. Moreover the highest weight vector of the corresponding Gaudin model is preserved. Based on these arguments Kulish postulated the Bethe vectors, the spectrum and the Bethe equations of the system.

Alternatively, the jordanian twist \cite{kulish5,Kulish2000cl,delolmo}
\begin{equation}
\label{Jtwist}
\mathcal{F}^{J}= e^{h\ox\sigma} = \exp({h\ox\frac{1}{2}\ln{(1 + 2\theta \xp)}})
\end{equation}
can be applied to the $\sl_{2}$-invariant spin system based on the Yang quantum R-matrix 

\begin{equation}
\label{RY}
R(\l;\eta)= I + \frac{\eta}{\l}\mathcal{P}
\end{equation}
where $\mathcal{P}$ is the permutation matrix in the tensor product $\bbC^{2}\ox\bbC^{2}$, to obtain the twisted $\sl_{2}$ spin system \cite{kulish5,Cirilo-Antoniozy}. The semiclassical limit $\eta\to 0$ of this system yields the same Gaudin model discussed by Kulish \cite{Kulishjs}. The twisted XXX spin chain related to the quantum R-matrix 
\begin{equation}
\label{Rtwisted}
R(\l;\eta;\theta) = R^{J}(\theta) + \frac{\eta}{\l}\mathcal{P},
\end{equation}
where $R^{J}(\theta) = \mathcal{F}^{J}_{21}(\mathcal{F}^{J}_{12})^{-1}$ is the $\sl_{2}$ jordanian quantum R-matrix studied in \cite{Gerstenhaber1992gt,Ogievetsky1994kr,Ballesteros1996eo}, whose homogeneous case was analyzed in \cite{kulish5}, will be discussed elsewhere \cite{Cirilo-Antoniozy}, here will only be presented the aspects relevant to the study of the corresponding Gaudin model. 

The $\mathfrak{L}$-operator of the quantum spin system on a one-dimensional lattice with $N$ sites coincides with $R$-matrix acting on a tensor product $V_{0} \otimes V_a$ of auxiliary space $V_{0}=\bbC^{2}$  and the space of states at site $a = 1, 2,\dots N$ 
\begin{equation}
\label{Lq}
\mathfrak{L}_{0a}(\l - z_a) = \left(\begin{array}{cc}e^{-\sigma_a} & \theta h_a e^{\sigma_a} \\0 & e^{\sigma_a}\end{array}\right) + \frac{\eta}{\l - z_{a}}\left(\begin{array}{cc}\frac{h_a}{2} & \xm_a \\\xp_a & -\frac{h_a}{2}\end{array}\right)\,, 
\end{equation}
where \(z_a\) is a parameter of inhomogeneity (site dependence). Corresponding monodromy matrix $\mathfrak{T}$ is an ordered product of the $\mathfrak{L}$-operators 
\begin{equation}
\label{T-q}
\mathfrak{T}(\l ; \{z_a\}_1^N) = \mathfrak{L}_{0N}(\l - z_N) \dots  
\mathfrak{L}_{01}(\l - z_1) = \underset{\longleftarrow}{\prod_{a=1}^{N}}
\mathfrak{L}_{0a} (\l - z_a) = \left(\begin{array}{cc}A(\l) & B(\l) \\C(\l) & D(\l)\end{array}\right)\,.  
\end{equation}
The commutation relations of the $\mathfrak{T}$-matrix entries follow form the FRT-relation
\begin{equation}
\label{RTT}
R_{12}(\l - \mu;\eta;\theta)\underset{1}{\mathfrak{T}}(\l)\underset{2}{\mathfrak{T}}(\mu) = \underset{2}{\mathfrak{T}}(\mu)\underset{1}{\mathfrak{T}}(\l)R_{12}(\l - \mu;\eta;\theta).  
\end{equation}
Multiplying \eqref{RTT} by $R_{12}^{-1}$ and taking the trace over $\bbC^{2} \ox \bbC^{2}$, one gets commutativity of the transfer matrix 
\begin{equation}
\label{tq}
\mathfrak{t}(\l) = \operatorname{tr} \; \mathfrak{T} (\l) =  A(\l) + D(\l)
\end{equation}
for different values of the spectral parameter $\mathfrak{t}(\l)\mathfrak{t}(\mu) = \mathfrak{t}(\mu)\mathfrak{t}(\l)$. 

It is of interest to choose different spins $l_a$ at different sites of the lattice, hence the following space of states 
$$
{\cal H} = \underset {a=1}{\overset{N}{\otimes}} V^{(l_a)}_a \;,
$$
with the highest spin vector $\O_{+}=\ox_{a=1}^{N}\o_{a}$. It is straightforward to show that 
\begin{equation}
\label{Ckill}
C(\l)\O_{+}=0,
\end{equation}
and
\begin{align}
\label{AD0eigen}
A(\l) \O_{+} = a(\l)\O_{+} =\prod_{a=1}^{N}\left(\frac{\l-z_{a}+\ell_{a}\eta/2}{\l-z_{a}}\right)\O_{+},\\
D(\l) \O_{+} = d(\l)\O_{+}=\prod_{a=1}^{N}\left(\frac{\l-z_{a}-\ell_{a}\eta/2}{\l-z_{a}}\right)\O_{+}.
\end{align}
As the first step in the application of the algebraic Bethe Ansatz to the twisted XXX spin chain one can confirm that the highest spin vector $\O_{+}$ is an eigenvector of the transfer matrix \eqref{tq}
\begin{equation}
\mathfrak{t}(\l)\O_{+} =\Lambda_{0}(\l)\O_{+},
\end{equation}
with $\Lambda_{0}(\l) = a(\l)+d(\l)$.
The next step is to show that $\Psi_{1}(\mu) = B(\mu)\O_{+}$ is also an eigenvector of the transfer matrix
\begin{align}
\mathfrak{t}(\l)\Psi_{1}(\mu) = \mathfrak{t}(\l)B(\mu)\O_{+} &= \Lambda_{1}(\l,\mu)\Psi_{1}(\mu) + \frac{\eta}{\l-\mu}(a(\mu)-d(\mu))\Psi_{1}(\l)\notag\\
& -  \te (a(\mu)-d(\mu))(a(\l)-d(\l))\O_{+}
\end{align}
where the eigenvalue $\Lambda_{1}(\l,\mu)$ and the Bethe equations are given by
\begin{align}
&\Lambda_{1}(\l,\mu) = \Lambda_{0}(\l) -  \frac{\eta}{\l-\mu}(a(\l)-d(\l)),\\
&a(\mu)-d(\mu)=0.
\end{align}
The following steps of the algebraic Bethe Ansatz are analogous having the same eigenvalues and Bethe equations as in the invariant case \cite{kulish5}, the details will be presented elsewhere \cite{Cirilo-Antoniozy}.

The corresponding Gaudin model  can be obtained from the spin system via semiclassical limit \cite{sklyanin4,lima} by setting $\theta=-\frac{\eta}{2}\xi$ and using the expansion in powers of $\eta$ of the monodromy matrix

\begin{equation}
\label{semiclassicallimit}
\mathfrak{T}(\l) = \operatorname{I} + \frac{\eta}{2} L(\l) + \mathcal{O}(\eta^{2}), 
\end{equation}
where 
\begin{equation}
\label{semiclassicalL}
L(\l)=\begin{pmatrix}
h(\l) & 2X^{-}(\l) \\
2X^{+}(\l) & - h(\l)
\end{pmatrix}
\end{equation}
and 
\begin{gather}
\label{semiclassicalgenerators}
h(\l) = \sum_{a=1}^{N} \left(\frac{h_{a}}{\l - z_a} + \xi \xp_{a}\right),\;
\xm(\l) = \sum_{a=1}^{N} \left(\frac{\xm_{a}}{\l - z_a} - \frac{\xi}{2} h_{a}\right),\; 
\xp(\l) = \sum_{a=1}^{N} \frac{\xp_{a}}{\l - z_a}.
\end{gather}
Substituting the expansion of the monodromy matrix \eqref{semiclassicallimit} and 
\begin{equation}
\label{semiclassicalR}
R(\l;\eta;\theta)\Big{|}_{\theta=-\frac{\eta}{2}\xi}=\operatorname{I} + \frac{\eta}{2} r_{\xi}(\l) + \mathcal{O}(\eta^{2}),
\end{equation}
with
\begin{equation}
\label{rclassico}
r_{\xi}(\l) =
\begin{pmatrix}
    \frac{1}{\l}	& \xi	& -\xi	& 0	\\
    0	& -\frac{1}{\l}	& \frac{2}{\l}	& \xi	\\
    0	& \frac{2}{\l}	& -\frac{1}{\l}	& -\xi	\\
    0  	& 0	& 0	& \frac{1}{\l}	
\end{pmatrix},
\end{equation}
into the FRT relations \eqref{RTT}, the first nontrivial term, the coeficient of $\eta^{2}$, is the so-called Sklyanin bracket \cite{sklyanin4}
\begin{equation}
\label{sklyaninbracket}
\left[\LI(\l),\LII(\mu)\right] = -\left[r_{\xi}(\l-\mu),\LI(\l) + \LII(\mu)\right].
\end{equation}
Moreover the central element $\Delta(\l)$ \cite{kulish10}, of the algebra \eqref{RTT}, admits the expansion in $\eta$ such that the second order term of $\mathfrak{t}(\l)-\Delta(\l)$ is the generating function of the integrals of motion of the corresponding Gaudin model \cite{sklyanin4} 
\begin{equation}
t(\l) = \frac{1}{2}\tr \;L^2(\l).
\end{equation}

The first step in the application of the algebraic Bethe Ansatz, to the Gaudin models, is to define appropriate creation operators that yield the Bethe states and consequently the spectrum of the generating function $t(\l)$. The creation operators used in the $\sl_{2}$-invariant Gaudin model coincide with one of the L-matrix entry \cite{gaudin2,sklyanin4}. However, in the present case these operators are not homogeneous polynomials of the generator $\xm(\l)$ and can be defined by the recursive relation as proposed by Kulish \cite{Kulishjs}. It is convenient to define $B^{(k)}_{M}(\mu_{1},\ldots,\mu_{M})$, a more general set of operators, in order to simplify the calculation of the commutators between the creation operators and the generating function of the integrals of motion $t(\l)$. These operators are symmetric functions of their arguments and they satisfy certain recursive relations. Their commutation relations with the generators of the loop algebra are straightforward to calculate and they are essential in the main step of the algebraic Bethe Ansatz. The creation operators of the $\sl_{2}$ Gaudin model with jordanian twist are the particular case $B_{M}(\mu_{1},\ldots,\mu_{M})=B^{(0)}_{M}(\mu_{1},\ldots,\mu_{M})$. Thus the commutation relations between the creation operators and $t(\l)$ are easily calculated. Therefore the corresponding Bethe vectors can be defined by the action of the creation operators on the highest spin vector $\O_{+}$ and thus the spectrum of the system determined. In this way the algebraic Bethe Ansatz is fully implemented confirming the result of Kulish \cite{Kulishjs} that the spectrum of the system coincides with the one of the $\sl_{2}$-invariant model and consequently the twisted Gaudin hamiltoneans have the same spectrum as in the $\sl_{2}$-invariant case, although the Bethe vectors of the two systems are different. However the Bethe vectors, in this case, are not eigenstates of the global Cartan generator $h_{gl}$.

Besides the problem of determining the spectrum of the system, via algebraic Bethe Ansatz, some properties of the creation operators are fundamental in calculating the inner products and the norms of the Bethe states. In particular, it turns out that the relation between the creation operators of the system and the ones in the untwisted case is essencial in determining the inner products of the Bethe vectors. In addition, it is necessary to consider the dual creation operators $B^{\ast}_{M}(\mu_{1},\ldots,\mu_{M})$ obtained by using the dual Gaudin model based on the following classical r-matrix 
\begin{equation}
\label{eq:a02}
r^{\ast}(\l)=\frac{1}{\l}(h\ox h + 2(\xp\ox\xm+\xm\ox\xp)) + \xi (h\ox \xm - \xm\ox h).
\end{equation}
As opposed to the $\sl_{2}$-invariant case, the Bethe vectors $\Psi_{M}(\mu_{1},\ldots,\mu_{M})$ are not orthogonal for different $M$'s. It should also be mentioned that, due to the jordanian twist, the dual generating function of integrals of motion $t^{*}(\l)$ is not equal to $t(\l)$, the generating function of integrals of motion of the original model, contrary to the $\sl_{2}$-invariant case, thus these operators are not hermitian. 

A connection between the Bethe vectors, when the Bethe equations are not imposed on their parameters, of the twisted Gaudin model to the solutions of the corresponding Knizhnik--Zamolodchikov equation, similarly to the $\sl_{2}$-invariant model, is based on some analytical properties of the creation operators $B_{M}(\mu_{1},\ldots,\mu_{M})$.

The article is organized as follows. In section 2 we discuss the Gaudin model based on the $\sl_{2}$ invariant r-matrix with jordanian twist emphasizing the creation operators $B_{M}$. Using the previously established properties of the creation operators $B_{M}$, in section 3 the spectrum and the Bethe vectors of the system are given. The dual creation operators are used to obtain the expressions for inner products and norms of Bethe states. In conclusions a relation between the Bethe vectors and the solutions to the Knizhnik-Zamolodchikov equation are discussed. In the appendix the proofs of the lemmas are given.

\section{$\sl_{2}$ Twisted Gaudin model}

The $\sl_{2}$-invariant classical r-matrix with jordanian twist \cite{drinfeld,Gerstenhaber1992gt, Ogievetsky1994kr} is the following element of $\sl_{2}\ox\sl_{2}$
\begin{equation}
\label{r-matrix}
r_{\xi}(\l) = \frac{c_{2}^{\ox}}{\l} + \xi r^{J} = \frac{1}{\l}(h\ox h + 2(\xp\ox\xm+\xm\ox\xp)) + \xi (h\ox \xp - \xp\ox h).
\end{equation}
The matrix form of $r_{\xi}(\l)$ in the fundamental representation of $\sl_{2}$ follows from \eqref{r-matrix} by replacing the appropriate matrices for the generators of $\sl_{2}$ is given explicitly by 
\begin{equation}
\label{eq:02}
r_{\xi}(\l) =
\begin{pmatrix}
    \frac{1}{\l}	& \xi	& -\xi	& 0	\\
    0	& -\frac{1}{\l}	& \frac{2}{\l}	& \xi	\\
    0	& \frac{2}{\l}	& -\frac{1}{\l}	& -\xi	\\
    0  	& 0	& 0	& \frac{1}{\l}	
\end{pmatrix},
\end{equation}
here $\l\in\bbC$ is the spectral and $\xi\in\bbC$ is the twisting parameter. Definition of the Gaudin model requires not only the classical r-matrix but also the L-operator
\begin{equation}
\label{eq:04}
L(\l) =
\begin{pmatrix}
h(\l) & 2X^{-}(\l) \\
2X^{+}(\l) & - h(\l)
\end{pmatrix}
\end{equation}
whose entries $\left(h(\l),X^{\pm}(\l))\right)$ are generators of the loop algebra $\Lxi (\sl_{2})$
defined by the Sklyanin linear bracket 
\begin{equation}
\label{linearbracket}
\left[\LI(\l),\LII(\mu)\right] = -\left[r_{12}(\l-\mu),\LI(\l) + \LII(\mu)\right].
\end{equation}
The corresponding commutation relations between the generators are explicitly given by 
\begin{align}
\left[h(\l),h(\mu)\right] &= 2\xi \left(\xp(\l) - \xp(\mu)\right)\notag\\
\left[\xm(\l),\xm(\mu)\right] &= -\xi \left(\xm(\l) - \xm(\mu) \right),\notag \\
\left[\xp(\l),\xm(\mu)\right] &= -\frac{h(\l) - h(\mu)}{\l - \mu} + \xi \xp(\l),\notag\\ 
\left[\xp(\l),\xp(\mu)\right] &= 0, 
\label{eq:05}
\\
\left[h(\l),\xm(\mu)\right] &= 2\frac{\xm(\l) - \xm(\mu)}{\l - \mu} + \xi h(\mu),\notag\\
\left[h(\l),\xp(\mu)\right] &= -2\frac{\xp(\l) - \xp(\mu)}{\l - \mu}.\notag
\end{align}
The usual $\sl_{2}$ loop algebra is recovered be setting $\xi=0$.

In order to define a dynamical system besides the algebra of observables $\Lxi (\sl_{2})$ an hamiltonian should be specified. Due to the Sklyanin linear bracket \eqref{linearbracket} the elements 
\begin{align}
\label{eq:10}
t(\l) = \frac{1}{2}\tr \;L^2(\l) &= h^{2}(\l) + 2(\xp(\l) \xm(\l) + \xm(\l) \xp(\l))\notag\\[0.3cm] 
& = h^{2}(\l) -2 h^{\prime}(\l) + 2\left(2\xm(\l)+\xi\right)\xp(\l)
\end{align}
generate an abelian subalgebra
\begin{equation}
\label{eq:11}
t(\l)t(\mu) = t(\mu)t(\l).
\end{equation}
One way to show \eqref{eq:11} is to notice that the commutation relation between $t(\l)$ and $L(\mu)$ can be written in the form
\begin{equation}
\label{eq:12}
\left[t(\l),L(\mu)\right] = \left[M(\l,\mu),L(\mu)\right],
\end{equation}
using (2.4-6), it is straightforward to calculate $M(\l,\mu)$
\begin{align}
\label{eq:12a}
M(\l,\mu) &= -\underset{1}{\tr}\left(r_{12}(\l - \mu)\LI(\l)\right)-\tfrac{1}{2}\underset{1}{\tr}\left(r_{12}^{2}(\l - \mu)\right) \notag\\[0.3cm]
    &=\begin{pmatrix}
    -2\tfrac{h(\l)}{\l - \mu} + 2\xi \xp(\l) & -4\tfrac{\xm(\l)}{\l - \mu} -2 \xi h(\l)\\
    -4\tfrac{\xp(\l)}{\l - \mu} & 2\tfrac{h(\l)}{\l - \mu} - 2\xi \xp(\l)\\
    \end{pmatrix}
- \tfrac{3}{(\l - \mu)^{2}}I_{2}.
\end{align}
Thus the commutators between the generating function $t(\l)$ and the generators of the loop algebra follow from (2.8-9) 
\begin{equation}
\label{eq:14}
\left[t(\l),\xp(\mu)\right] = 4\frac{\xp(\mu)h(\l) - \xp(\l)h(\mu)}{\l - \mu} - 4\,\xi\, \xp(\l)\xp(\mu),
\end{equation}

\begin{equation}
\label{eq:15}
\left[t(\l),h(\mu)\right] = 8\frac{\xm(\mu)\xp(\l) - \xm(\l)\xp(\mu)}{\l - \mu} - 4\,\xi\, h(\l)\xp(\mu),
\end{equation}

\begin{equation}
\label{eq:16}
\begin{split}
\left[t(\l),\xm(\mu)\right] &= 4\frac{\xm(\l)h(\mu) - \xm(\mu)h(\l)}{\l - \mu} + 2 \xi h(\l)h(\mu)\\
&\quad + 4\xi\xm(\mu)\xp(\l) + 2\xi^{2}\xp(\mu).
\end{split}
\end{equation}

Preserving some generality, the representation space $\cH$ of the Sklyanin algebra (2.4-5) can be a highest spin $\rho(\l)$ representation with the highest spin vector $\O_{+}$
\begin{equation}
\label{vacuum}
\xp(\l)\O_{+}=0, \quad h(\l)\O_{+}=\rho(\l)\O_{+}.
\end{equation}
The spectrum and the eigenstates of $t(\l)$ can be studied in this general setting. However to have a physical interpretation a local realization 
\begin{equation}
\label{eq:07}
\cH = V_1\ox\cdots\ox V_N.
\end{equation}
as tensor product of $\sl_{2}$-modules is needed. Then the L-operator is given by
\begin{equation}
\label{eq:08}
L(\l) = \sum_{a=1}^{N} \left(\frac{1}{\l - z_a} 
\begin{pmatrix}
h_{a} & 2\xm_{a} \\
2\xp_{a} & - h_{a}
\end{pmatrix}
+ \xi
\begin{pmatrix}
\xp_{a} & -h_{a} \\
0 & - \xp_{a}
\end{pmatrix}
\right),
\end{equation}
were $Y_{a} = (h_{a},X^{\pm}_{a})\in \operatorname{End}(V_{a})$ are $\sl_{2}$ generators in representation $V_{a}$, associated with each site $a$. For convenience, the generators $Y(\l)=(h(\l),X^{\pm}(\l))$ are written down explicitly
\begin{gather}
\label{eq:09}
h(\l) = \sum_{a=1}^{N} \left(\frac{h_{a}}{\l - z_a} + \xi \xp_{a}\right),\;
\xm(\l) = \sum_{a=1}^{N} \left(\frac{\xm_{a}}{\l - z_a} - \frac{\xi}{2} h_{a}\right),\; 
\xp(\l) = \sum_{a=1}^{N} \frac{\xp_{a}}{\l - z_a}.
\end{gather}
In this realization it is useful to consider the expressions of the generators of the global $\sl_{2}$ Lie algebra in terms of the local ones
\begin{equation}
\label{globais}
Y_{gl}= \sum_{a=1}^{N}Y_{a},
\end{equation}
where $Y = (h,X^{\pm})$. Also, the following notation is useful  
\begin{equation}
\label{untwisted}
Y(\l)_{0}= Y(\l)|_{\xi=0},
\end{equation}
then $h(\l)=h(\l)_{0}+ \xi \xp_{gl}$.

A representation of the model with the above Gaudin realization is obtained by considering at each site $a$ an irreducible representation $V_{a}^{(\ell_{a})}$ of $\sl_{2}$ with highest weight $\ell_{a}$ corresponding to a singular vector $\o_{a}\in V_{a}^{(\ell_{a})} \text{ such that } \xp_{a}\o_{a}=0 \text{ and } h_{a}\o_{a}=\ell_{a}\,\o_{a}$. Thus the space of states is
\begin{equation}
\label{eq:17}
\cH = V_{1}^{(\ell_{1})}\ox\cdots\ox V_{N}^{(\ell_{N})},
\end{equation}
with the highest spin vector \eqref{vacuum}
\begin{equation}
\label{Omega}
\O_{+}=\o_{1}\ox\cdots\ox\o_{N}
\end{equation}
and the corresponding highest spin
\begin{equation}
\label{Ro}
\rho(\l)=\sum_{a=1}^{N}\frac{\ell_{a}}{\l-z_{a}}.
\end{equation}

The Gaudin hamiltoneans 
\begin{equation}
\label{Ghamiltoneans}
H^{(a)} = \sum_{b\neq a}^{N} r_{ab}(z_{a}-z_{b})=  \sum_{b\neq a}^{N}\left(\frac{c_{2}^{\,\ox}(a,b)}{z_a - z_b} + \xi \left(h_{a} X_{b}^{+} - X_{a}^{+}h_{b} \right)\right),
\end{equation}
where $c_{2}^{\,\ox}(a,b) = h_{a}\ox h_{b} + 2(\xp_{a}\ox\xm_{b} + \xm_{a}\ox\xp_{b})$, can be obtained as the residues of the generating function $t(\l)$ at the points $\l=z_{a}$, $a=1,\ldots,N$ using the expansion
\begin{equation}
\label{poleexp}
t(\l) = \sum_{a=1}^{N}\left(\frac{c_{2}(a)}{(\l - z_a)^{2}} + \frac{2H^{(a)}}{\l - z_a} \right) + \xi^{2} \sum_{a,b=1}^{N}\xp_{a}\xp_{b},
\end{equation}
here $c_{2}(a)= h_{a}^{2} + 2h_{a} + 4\xm_{a}\xp_{a}$ is the $\sl_{2}$ Casimir at site $a$. As opposed to the $\sl_{2}$-invariant case, the generating function \eqref{poleexp} commutes only with the generator $\xp_{gl}$. 

The first step in the algebraic Bethe Ansatz is to define appropriate creation operators that yield the Bethe states and consequently the spectrum of the generating function $t(\l)$. The creation operators used in the $\sl_{2}$-invariant Gaudin model coincide with one of the L-matrix entry \cite{gaudin2,sklyanin4}. However, in the present case these operators are not homogeneous polynomials of the generator $\xm(\l)$. It is convenient to define a more general set of operators in order to simplify the presentation.
%
%
\begin{definition}
\label{def:1}
Given two integers $M$ and $k\geq 0$ consider the operators
\begin{align}
\label{BMk}
B^{(k)}_{M}(\mu_{1},\ldots,\mu_{M}) &=\left(\xm(\mu_{1})+k\xi\right)\left(\xm(\mu_{2})+(k+1)\xi\right)\cdots\left(\xm(\mu_{M}) + (M+k-1)\xi\right)\notag\\
&= \prod_{\substack{n=k\\ \rightarrow}}^{M+k-1} \left(\xm(\mu_{n-k+1}) + n\,\xi\right),
\end{align}
with $B^{(k)}_{0}=1$ and $B^{(k)}_{M}=0$ for $M<0$.
\end{definition}
The following lemma describes some properties of the $B^{(k)}_{M}(\mu_{1},\ldots,\mu_{M})$ operators which will be used later on.
%

\begin{lemma}
\label{lem:1}
Some useful properties of $B^{(k)}_{M}(\mu_{1},\ldots,\mu_{M})$ operators are
\begin{description}
\item[i.]
The operators $B^{(k)}_{M}(\mu_{1},\ldots,\mu_{M})$ are symmetric functions of their arguments.
\item[ii.]
$B^{(k)}_{M}(\mu_{1},\ldots,\mu_{M}) = B^{(k)}_{M-1}(\mu_{1},\ldots,\mu_{M-1})\left(B_{1}^{(0)}(\mu_{M}) + (M+k-1) \xi\right)$.
\item[iii.]
$B^{(k)}_{M}(\bmu)= B^{(k-1)}_{M}(\bmu) + \xi\sum_{i=1}^{M}B^{(k)}_{M-1}(\bmu^{(i)})$, here $\bmu$ is a set of $M=|\bmu|$ complex scalars with a particular ordering $\bmu=\left\{\mu_{1},\ldots,\mu_{M}\right\}$ assumed for convenience and the notation
$$
\bmu^{(i_{1},\ldots,i_{k})}=\bmu\setminus\left\{\mu_{i_{1}},\ldots,\mu_{i_{k}}\right\}
$$
for any distinct $i_{1},\ldots,i_{k}\in\left\{1,\ldots,M\right\}$.
\item[iv.]
$B^{(k)}_{M}(\bmu)= B^{(k)}_{1}(\mu_{1})B^{(k+1)}_{M-1}(\bmu^{(1)})$.
\end{description}
\end{lemma}

Implementation of the algebraic Bethe Ansatz requires the commutation relations\break between the generators of the loop algebra $\Lxi(\sl_{2})$ and the $B^{(k)}_{M}(\mu_{1},\ldots,\mu_{M})$ operators. To this end we first introduce the notation for the Bethe operators
\begin{equation}
\label{eq:26}
\betheop_{M} (\l;\bmu) = h(\l) + \sum_{\mu\in\bmu}\frac{2}{\mu - \l},
\end{equation}
where $\bmu=\{\mu_{1},\ldots,\mu_{M-1}\}$ and $\l\in\bbC\setminus\bmu$. In the particular case $M=1$, $\betheop_{1} (\l;\emptyset) = h(\l)$ and will be denoted by $\betheop_{1} (\l)$. The required commutators are given in the following lemma.

\begin{lemma}
\label{lem:2}
The commutation relation between the generators $h(\l)$, $X^{\pm}(\l)$ and the \break $B^{(k)}_{M}(\mu_{1},\ldots,\mu_{M})$ operators are given by
\begin{align}
\label{eq:27}
h(\l)B^{(k)}_{M}(\bmu) &= B^{(k)}_{M}(\bmu) h(\l) + 2\sum_{i=1}^{M}\frac{B^{(k)}_{M}(\l\cup\bmu^{(i)})-B^{(k)}_{M}(\bmu)}{\l-\mu_{i}}\notag\\
&\quad + \xi \sum_{i=1}^{M}B^{(k+1)}_{M-1}(\bmu^{(i)})\betheop_{M}(\mu_{i};\bmu^{(i)});\\
%
\label{eq:28}
\xp(\l)B^{(k)}_{M}(\bmu) &= B^{(k)}_{M}(\bmu) \xp(\l) - \sum_{i=1}^{M} B^{(k+1)}_{M-1}(\bmu^{(i)}) \frac{\betheop_{M}(\l;\bmu^{(i)})-\betheop_{M}(\mu_{i};\bmu^{(i)})}{\l-\mu_{i}} \notag\\
&\quad - 2\sum_{\substack{i,j=1\\i<j}}^{M}\frac{B^{(k+1)}_{M-1}(\{\l\}\cup\bmu^{(i,j)})}{(\l-\mu_{i})(\l-\mu_{j})} + \xi \sum_{i=1}^{M} B^{(k+1)}_{M-1}(\bmu^{(i)})\xp(\l);\\
%
\label{eq:29}
\xm(\l)B^{(k)}_{M}(\bmu) &= B^{(k)}_{M}(\bmu)\left(\xm(\l) + M\xi \right) -\xi \sum_{i=1}^{M} B^{(k)}_{M}(\l\cup\bmu^{(i)})\notag\\
&=B^{(k)}_{M+1}(\l\cup\bmu) -\xi \sum_{i=1}^{M} B^{(k)}_{M}(\l\cup\bmu^{(i)}).
\end{align}
\end{lemma}

The $B$-operators that define the Bethe states of the system were proposed by\break Kulish \cite{Kulishjs}. These operators are the particular case $k=0$ of \eqref{BMk} and will be denoted by $B_{M}(\mu_{1},\ldots,\mu_{M})=B^{(0)}_{M}(\mu_{1},\ldots,\mu_{M})$. A recursive relation of the B-operators follows from ii. of the lemma \ref{lem:1}
\begin{equation}
\label{eq:22}
B_{M}(\bmu) = B_{M-1}(\bmu^{(M)})\left(\xm(\mu_{M}) + (M-1)\xi \right).
\end{equation}
It may be useful to write down explicitly first few B-operators
\begin{align*}
B_{0}=1,\quad B_{1}(\mu) &= \xm(\mu),\quad 
B_{2}(\mu_{1},\mu_{2}) = \xm(\mu_{1} )\xm(\mu_{2} ) + \xi \xm(\mu_{1} )\\[0.3cm]
B_{3}(\mu_{1},\mu_{2},\mu_{3}) &= \xm(\mu_{1} )\xm(\mu_{2} )\xm(\mu_{3} ) + 2\xi \xm(\mu_{1} )\xm(\mu_{2} )\\
& \quad + \xi \xm(\mu_{1} )\xm(\mu_{3} ) + 2\xi^{2}\xm(\mu_{1} ),
\end{align*}
As a particular case of lemma \ref{lem:2} the commutation relation between loop algebra generators $h(\l)$, $X^{\pm}(\l)$ and the B-operators are given by settting $k=0$ in (2.26-28)
\begin{equation}
\label{eq:27a}
h(\l)B_{M}(\bmu) = B_{M}(\bmu) h(\l) + \sum_{i=1}^{M}\left(2\frac{B_{M}(\l\cup\bmu^{(i)})-B_{M}(\bmu)}{\l-\mu_{i}} +\xi B^{(1)}_{M-1}(\bmu^{(i)})\betheop_{M}(\mu_{i};\bmu^{(i)})\right),
\end{equation}
\begin{align}
\label{eq:28a}
\xp(\l)B_{M}(\bmu) &= B_{M}(\bmu) \xp(\l) - \sum_{i=1}^{M} B^{(1)}_{M-1}(\bmu^{(i)})\left( \frac{\betheop_{M}(\l;\bmu^{(i)})-\betheop_{M}(\mu_{i};\bmu^{(i)})}{\l-\mu_{i}}-\xi\xp(\l)\right)\notag\\
&\quad - 2\sum_{\substack{i,j=1\\i<j}}^{M}\frac{B^{(1)}_{M-1}(\{\l\}\cup\bmu^{(i,j)})}{(\l-\mu_{i})(\l-\mu_{j})},
\end{align}
\begin{align}
\label{eq:29a}
\xm(\l)B_{M}(\bmu) &= B_{M}(\bmu)\left(\xm(\l) + M\xi \right) -\xi \sum_{i=1}^{M} B_{M}(\l\cup\bmu^{(i)}).
\end{align}

The crucial step in the algebraic Bethe Ansatz is to determine the commutation relations between the creation operators, in this case the B-operators, and the generating function $t(\l)$. The main lemma, based on the results established previously in this section, will given a complete expression for the required commutator.
\begin{lemma}
\label{lem:4}
The generating function $t(\l)$ has the following commutation relation with the $B$-operators:
\begin{equation}
\label{eq:32}
\begin{split}
t(\l)B_{M}(\bmu) &= B_{M}(\bmu)  \left( t(\l) - \sum_{i=1}^{M} \frac{4h(\l)}{\l - \mu_{i}} + \sum_{i<j}^{M} \frac{8}{\left(\l - \mu_{i}\right) \left(\l - \mu_{j}\right)}\right)\\
& + 4\sum_{i=1}^{M} \frac{B_{M}(\l\cup\bmu^{(i)})}{\l - \mu_{i}}\betheop_{M}(\mu_{i};\bmu^{(i)})+2\xi\sum_{i=1}^{M}B^{(1)}_{M-1}(\bmu^{(i)})h(\l) \betheop_{M}(\mu_{i};\bmu^{(i)})\\ 
& + 4\xi\sum_{\substack{i,j=1\\i\neq j}}^{M} \frac{B^{(1)}_{M-1}(\l\cup\bmu^{(i,j)})-B^{(1)}_{M-1}(\bmu^{(i)})}{\l-\mu_{j}}\betheop_{M}(\mu_{i};\bmu^{(i)})\\
& + \xi^{2}\sum_{\substack{i,j=1\\i\neq j}}^{M} B^{(2)}_{M-2}(\bmu^{(i,j)})\betheop_{M-1}(\mu_{j};\bmu^{(i,j)})\betheop_{M}(\mu_{i};\bmu^{(i)})\\
& + 4 M \xi B_{M}(\bmu)\xp(\l) + 2\xi^{2}\sum_{i=1}^{M} B^{(1)}_{M-1}(\bmu^{(i)}) \xp(\mu_{i}).
\end{split}
\end{equation}
\end{lemma}
\begin{proof}
The case $M=1$ can be obtained directly from the M-matrix \eqref{eq:12a} and is given by \eqref{eq:16}. For $M > 1$ the commutator between the operator $t(\l)$ and the corresponding B-operator is to be calculated directly using expression \eqref{eq:10}
\begin{equation*}
\begin{split}
\left[t(\l),B_{M}(\bmu)\right] &= \left[h(\l),\left[h(\l),B_{M}(\bmu)\right]\right] + 2\left[h(\l),B_{M}(\bmu)\right] h(\l) - 2 \frac{d}{d\l} \left[h(\l),B_{M}(\bmu)\right]\\
&\quad + 2\left(2\xm(\l)+\xi\right)\left[\xp(\l),B_{M}(\bmu)\right] +4\left[\xm(\l),B_{M}(\bmu)\right]\xp(\l).
\end{split}
\end{equation*}
The terms in the previous expression only involve the commutators between the generators of the loop algebra and the B-operators. Each term in the above equation will be calculated separately. Using the commutators \eqref{eq:27} and (2.30-32) the first term is given by
\begin{align*}
&\left[h(\l),\left[h(\l),B_{M}(\bmu)\right]\right] =B_{M}(\bmu)\sum_{i<j}^{M}\frac{8}{(\l-\mu_{i})(\l-\mu_{j})}\\
& - 4\sum_{i=1}^{M}\left(\frac{B_{M}(\{\l\}\cup\bmu^{(i)}) - B_{M}(\bmu)}{(\l-\mu_{i})^{2}} - \frac{\frac{d}{d\l}B_{M}(\{\l\}\cup\bmu^{(i)})}{\l-\mu_{i}}\right)\\
& + 8\sum_{i < j}^{M}\frac{\xm(\l)B^{(1)}_{M-1}(\{\l\}\cup\bmu^{(i,j)})}{(\l-\mu_{i})(\l-\mu_{j})} - 4\sum_{i=1}^{M}\frac{B_{M}(\{\l\}\cup\bmu^{(i)})}{\l-\mu_{i}}\sum_{j\neq i}^{M}\frac{2}{\l-\mu_{j}}\\
& + 4\xi\sum_{i<j}^{M}\frac{B^{(1)}_{M-1}(\{\l\}\cup\bmu^{(i,j)})}{(\l-\mu_{i})(\l-\mu_{j})} + 2\xi\sum_{i=1}^{M} B^{(1)}_{M-1}(\bmu^{(i)})\frac{\betheop_{M}(\l;\bmu^{(i)})-\betheop_{M}(\mu_{i}\bmu^{(i)})}{\l-\mu_{i}}\\
+ \xi\sum_{i\neq j}^{M} &\left(4\frac{B^{(1)}_{M-1}(\l\cup\bmu^{(i,j)})-B^{(1)}_{M-1}(\bmu^{(i)})}{\l-\mu_{j}} + \xi B^{(2)}_{M-2}(\bmu^{(i,j)})\betheop_{M-1}(\mu_{j};\bmu^{(i,j)})\right)\betheop_{M}(\mu_{i};\bmu^{(i)})\\
& + 2\xi^{2}\sum_{i=1}^{M} B^{(1)}_{M-1}(\bmu^{(i)})(\xp(\l)-\xp(\mu_{i}))
\end{align*}
The commutator between the generator $h(\l)$ and $B_{M}(\bmu)$ is also used to determine the next two terms
\begin{equation*}
\begin{split}
&2\left[h(\l),B_{M}(\bmu)\right] h(\l) = B_{M}(\bmu)\sum_{i=1}^{M}\frac{-4\,h(\l)}{\l-\mu_{i}} + 4\sum_{i=1}^{M}\frac{B_{M}(\{\l\}\cup\bmu^{(i)})}{\l-\mu_{i}}h(\l)\\
& + 2\xi\sum_{i=1}^{M}B_{M-1}^{(1)}(\bmu^{(i)}) h(\l)\betheop_{M}(\mu_{i};\bmu^{(i)})-4\xi^{2}\sum_{i=1}^{M}B_{M-1}^{(1)}(\bmu^{(i)})(\xp(\l)-\xp(\mu_{i})),\\
\intertext{and after diferentiating \eqref{eq:27a}}
&- 2 \frac{d}{d\l} \left[h(\l),B_{M}(\bmu)\right] = 4\sum_{i=1}^{M}\frac{B_{M}(\{\l\}\cup\bmu^{(i)}) - B_{M}(\bmu)}{(\l-\mu_{i})^{2}}-\frac{\frac{d}{d\l}B_{M}(\{\l\}\cup\bmu^{(i)})}{\l-\mu_{i}}.\\
\end{split}
\end{equation*}
The next term follows directly from \eqref{eq:28a}
\begin{align*}
& 2\left(2\xm(\l)+\xi\right)\left[\xp(\l),B_{M}(\bmu)\right] = -4\sum_{i=1}^{M} B_{M} (\{\l\}\cup\bmu^{(i)}) \frac{\betheop_{M}(\l;\bmu^{(i)})-\betheop_{M}(\mu_{i};\bmu^{(i)})}{\l-\mu_{i}}\\
& -8\sum_{i<j}^{M}\frac{\xm(\l)B^{(1)}_{M-1}(\{\l\}\cup\bmu^{(i,j)})}{(\l-\mu_{i})(\l-\mu_{j})} - 2 \xi\sum_{i=1}^{M} B^{(1)}_{M-1}(\bmu^{(i)})\frac{\betheop_{M}(\l;\bmu^{(i)})-\betheop_{M}(\mu_{i};\bmu^{(i)})}{\l-\mu_{i}}\\
& -4\xi\sum_{i<j}^{M}\frac{B^{(1)}_{M-1}(\{\l\}\cup\bmu^{(i,j)})}{(\l-\mu_{i})(\l-\mu_{j})} + 4\xi\,\sum_{i=1}^{M} B_{M}(\{\l\}\cup\bmu^{(i)})\xp(\l)\\
&+2\xi^{2}\,\sum_{i=1}^{M} B^{(1)}_{M-1}(\bmu^{(i)})\xp(\l).
\intertext{The last term is obtained from \eqref{eq:29a}}
& 4\left[\xm(\l),B_{M}(\bmu)\right]\xp(\l) = 4\,M\,\xi\, B_{M}(\bmu)\xp(\l) - 4\,\xi\sum_{i=1}^{M}B_{M}(\{\l\}\cup\bmu^{(i)})\xp(\l).
\end{align*}
When all terms in $\left[t(\l),B_{M}(\bmu)\right]$ are put together it is straightforward to obtain the formula \eqref{eq:32}.
\end{proof}

Besides the problem of determining the spectrum of the system, via algebraic Bethe Ansatz, some properties of the B-operators are fundamental in calculating the inner products and the norms of the Bethe states. To this end the relation between the B-operators and the untwisted ones is established  using the Gaudin realization \eqref{eq:09}:
\begin{equation}
\label{eq:54}
B_{M}(\bmu) = \sum_{k=0}^{M-1}\xi^{k} \sum_{j_{1<\cdots<j_{M-k}}}^{M} B_{M-k}(\mu_{j_{1}},\ldots,\mu_{j_{M-k}})_{0} \;\pop_{k}^{(M-k)} + \xi^{M}\pop_{M}
\end{equation}
where $$B_{M-k}(\mu_{j_{1}},\ldots,\mu_{j_{M-k}})_{0}=B_{M-k}(\mu_{j_{1}},\ldots,\mu_{j_{M-k}})|_{\xi=0}$$ and also
\begin{equation}
\label{phat}
\pop_{i}^{(j)}=(-\frac{h_{gl}}{2}+j)\cdots(-\frac{h_{gl}}{2}+i+j-1)\quad\text{with}\quad \pop_{i}= \pop_{i}^{(0)},
\end{equation}
are operators defined for any integers $i$ and $j$. 

Moreover, some analytical properties of the B-operators are important when solving the corresponding  Knizhnik-Zamolodchikov equation.
\begin{lemma}
\label{lem:2a}
The $B$-operators satisfy the following identity
\begin{equation}
\label{eq:2a}
\frac{\partial}{\partial z_{a}} B_{M}(\bmu) = -\sum_{i=1}^{M}\frac{\partial}{\partial \mu_{i}}\left(\xm_{a}(\mu_{i})B^{(1)}_{M-1}(\bmu^{(i)})\right)
\end{equation}
where $\xm_{a}(\mu_{i})=\frac{\xm_{a}}{\mu_{i} - z_{a}}-\frac{\xi}{2}h_{a}$ for site $a=1,\ldots,N$.
\end{lemma}
Using the relevant properties of the B-operators established in this section, the fundamental description of the $\sl_{2}$ Gaudin model with jordanian twist can be obtained.

\section{Spectrum and Bethe vectors of the twisted $\sl_{2}$ Gaudin model}

In this section the spectrum and the Bethe vectors of the twisted $\sl_{2}$ Gaudin model as well as their inner products and norms will be determined by applying the algebraic Bethe Ansatz. The first step is to define the Bethe vectors $\Psi_{M}(\bmu)=B_{M}(\bmu)\O_{+}$ by the action of the $B$-operators on the highest spin vector  $\O_{+}$. Then the key observation is that highest spin vector $\O_{+}$ is an eigenvector of the generating function of integrals of motion $t(\l)$. Finally the spectrum of the system is obtained as a consequence of the commutation relations between $t(\l)$ and $B_{M}(\bmu)$ lemma \ref{lem:4}, \eqref{eq:32}. The unwanted terms coming from the commutator are annihilated by the Bethe equations on the parameters $\bmu=\{\mu_{1},\ldots,\mu_{M}\}$ as well as by the condition $\xp(\l)\O_{+}=0$. Hence the algebraic Bethe Ansatz can be resumed in the following theorem.
%

\begin{theorem}
\label{thm:1}
The highest weight vector $\O_{+}$ is an eigenvector of $t(\l)$ 
\begin{equation}
\label{Lambda0}
t(\l)\O_{+} = \L_{0}(\l) \O_{+}
\end{equation}
with the corresponding eigenvalue 
\begin{equation}
\label{eq:35}
\L_{0}(\l) = \rho^{2}(\l) - 2\rho^{\prime}(\l) 
= \sum_{a=1}^{N}\frac{2}{\l - z_{a}}\left(\sum_{b\neq a}^{N}\frac{\ell_{a}\ell_{b}}{z_{a} - z_{b}}\right)
+ \sum_{a=1}^{N}\frac{\ell_{a}(\ell_{a}+2)}{(\l - z_{a})^{2}} .
\end{equation}
Furthermore, the action of the $B$-operators on the highest spin vector $\O_{+}$ yields the 
Bethe vectors
\begin{equation}
\label{eq:36}
\Psi_{M}(\bmu)=B_{M}(\bmu)\O_{+},
\end{equation}
such that
\begin{equation}
\label{eq:37}
t(\l)\Psi_{M}(\bmu) = \L_{M}(\l;\bmu)\Psi_{M}(\bmu),
\end{equation}
with the eigenvalues
\begin{equation}
\label{eq:38}
\L_{M}(\l;\bmu) = \rho_{M}^{2}(\l;\bmu) - 2\frac{\partial \rho_{M}}{\partial\l}(\l;\bmu)\quad\text{and}\quad\rho_{M}(\l;\bmu)=\rho(\l)-\sum_{i=1}^{M}\frac{2}{\l-\mu_{i}},
\end{equation}
provided that the Bethe equations are imposed on the parameters $\bmu=\left\{\mu_{1},\ldots,\mu_{M}\right\}$
\begin{equation}
\label{eq:39}
\sum_{a=1}^{N}\frac{\ell_{a}}{\mu_{i}-z_{a}} - \sum_{j\neq i}^{M}\frac{2}{\mu_{i}-\mu_{j}} = 0,\quad i=1,\dots,M.
\end{equation}
\end{theorem}
\begin{proof}
A consequence of \eqref{eq:10}, \eqref{vacuum} and (2.20-21) is
$$
t(\l)\O_{+}=(h^{2}(\l) -2 h^{\prime}(\l))\O_{+}=(\rho^{2}(\l) -2 \rho^{\prime}(\l))\O_{+}=\L_{0}(\l)\O_{+}.
$$
The action of $t(\l)$ on the Bethe vectors $\Psi_{M}(\bmu)$ \eqref{eq:36} is evident from lemma \ref{lem:4}
\begin{align}
\label{tcomPsi}
& t(\l)\Psi_{M}(\bmu) = t(\l) B_{M}(\bmu)\O_{+} = \L_{0}(\l)\Psi_{M}(\bmu)+\left[t(\l),B_{M}(\bmu)\right]\O_{+}\notag\\
&= \left( \L_{0}(\l) - \sum_{i=1}^{M} \frac{4\rho(\l)}{\l - \mu_{i}} + \sum_{i<j}^{M} \frac{8}{(\l - \mu_{i})(\l - \mu_{j})}\right)\Psi_{M}(\bmu) + 4\sum_{i=1}^{M} \frac{\Psi_{M}(\l\cup\bmu^{(i)})}{\l - \mu_{i}}\rho_{M}(\mu_{i};\bmu^{(i)})\notag\\
& + 2\xi\sum_{i=1}^{M}\left( \rho(\l) \Psi^{(1)}_{M-1}(\bmu^{(i)}) + 2 \sum_{i\neq j}^{M} \frac{\Psi^{(1)}_{M-1}(\l\cup\bmu^{(i,j)})-\Psi^{(1)}_{M-1}(\bmu^{(i)})}{\l-\mu_{j}}\right)\rho_{M}(\mu_{i};\bmu^{(i)})\notag\\ 
& + \xi^{2} \sum_{i=1}^{M}\left(\sum_{j\neq i}^{M} \rho_{M-1}(\mu_{j};\bmu^{(i,j)}) \Psi^{(2)}_{M-2}(\bmu^{(i,j)})\right)\rho_{M}(\mu_{i};\bmu^{(i)}),
\end{align}
where $\Psi^{(k)}_{M}(\bmu)=B^{(k)}_{M}(\bmu)\O_{+}$ and $\xp(\l)\O_{+}=0$. When the Bethe equations are imposed on the parameters $\bmu=\{\mu_{1},\ldots,\mu_{M}\}$
\begin{equation}
\rho_{M}(\mu_{i};\bmu^{(i)})=\sum_{a=1}^{N}\frac{\ell_{a}}{\mu_{i}-z_{a}} - \sum_{j\neq i}^{M}\frac{2}{\mu_{i}-\mu_{j}} =0,\quad i=1,\ldots,M,
\end{equation}
the unwanted terms in \eqref{tcomPsi} vanish and hence $t(\l)\Psi_{M}(\bmu) = \L_{M}(\l;\bmu)\Psi_{M}(\bmu)$ with 
\begin{equation}
\L_{M}(\l;\bmu)=\L_{0}(\l)- \sum_{i=1}^{M} \frac{4\rho(\l)}{\l - \mu_{i}} + \sum_{i<j}^{M} \frac{8}{(\l - \mu_{i})(\l - \mu_{j})}=\rho_{M}^{2}(\l;\bmu) - 2\frac{\partial \rho_{M}}{\partial\l}(\l;\bmu).
\end{equation}
\end{proof}

Due to the pole expansion \eqref{poleexp} of $t(\l)$ the previous theorem yields the spectrum of the Gaudin hamiltoneans.
\begin{corollary}
\label{cor:2}
The Bethe vectors $\Psi_{M}(\bmu)$ are eigenvectors of the Gaudin hamiltonians \eqref{Ghamiltoneans}
\begin{equation}
\label{GMspc}
H^{(a)}\Psi_{M}(\bmu) = E_{M}^{(a)}\Psi_{M}(\bmu)
\end{equation}
with the eigenvalues
\begin{equation}
\label{GMVp}
E_{M}^{(a)}=\sum_{b\neq a}^{N}\frac{\ell_{a}\ell_{b}}{z_{a}-z_{b}}-\sum_{i=1}^{M}\frac{2\ell_{a}}{z_{a}-\mu_{i}}.
\end{equation}
\end{corollary}

\begin{remark}
The fact that the spectrum of the $\sl_{2}$-invariant Gaudin model remains is not affected by the jordanian twist can also be deduced from the following expression
\begin{equation}
t(\l) = t(\l)_{0} +2\xi\left(h(\l)_{0}\xp_{gl} + 2\, \hat{p}_{1}^{(1)}\xp(\l)\right) +\xi^{2} (\xp_{gl})^{2}.
\end{equation}

\end{remark}

\begin{remark}
\label{rem:3}
From the Gaudin realization \eqref{eq:09} follows that $\xi\,\xp_{gl} = \lim_{\l\rightarrow\infty} h(\l)$ and together with \eqref{eq:27a} is straightforward to obtain the action of the global generator $\xp_{gl}$ on the Bethe vectors
\begin{equation}
\label{eq:40}
\xp_{gl}\Psi_{M}(\bmu) = \sum_{i=1}^{M}\b_{M}(\mu_{i};\bmu^{(i)}) \Psi^{(1)}_{M-1}(\bmu^{(i)}).
\end{equation}
Thus the Bethe vectors are annihilated by the global generator $\xp_{gl}$
\begin{equation}
\label{eq:41}
\xp_{gl}\Psi_{M}(\bmu) = 0
\end{equation}
once the Bethe equations \eqref{eq:38} are imposed.
\end{remark}
\begin{remark}
Analogously from \eqref{eq:09} follows that $-\frac{\xi}{2}\,h_{gl} = \lim_{\l\rightarrow\infty} \xm(\l)$ and with \eqref{eq:29a} and \textbf{ii.} of lemma \ref{lem:1} shows that Bethe vectors are not eigenstates of the global generator $h_{gl}$,
\begin{equation}
\label{eq:53}
\begin{split}
h_{gl}\Psi_{M}(\bmu) &= -2\, p_{1}^{(M)}\Psi_{M}(\bmu) + 2\, \xi\, p_{1}^{(M-1)}\sum_{i=1}^{M}\Psi_{M-1}(\bmu^{(i)}),
\end{split}\end{equation}
where $p = -\frac{1}{2}\sum_{a=1}^{N}\ell_{a}$ and $p^{(k)}_{l}=(p+k)_{l} = (p+k)\cdots(p+l+k-1)$ is the Pochhammer symbol.
\end{remark}

As it was shown already, the jordanian twist of the $\sl_{2}$-invariant Gaudin model preserves its spectrum but the Bethe vectors are different, thus their inner products and norms are changed also. In order to determine the inner products and norms of the Bethe vectors it is of interest to consider the dual B-operators, obtained by using the dual
\begin{gather}
\label{eq:a05}
\xmd(\l) = \sum_{a=1}^{N} \left(\frac{\xp_{a}}{\l - z_a} - \frac{\xi}{2} h_{a}\right),
\end{gather}
explicitly given by
\begin{equation}
\label{eq:a09}
B^{*}_{M}(\l_{1},\ldots,\l_{M}) = \prod_{\substack{k=1\\\leftarrow}}^{M} (\xmd(\l_{k}) + (k-1)\xi).
\end{equation}
As opposed to the $\sl_{2}$-invariant case here evidently the dual B-operators do not annihilate the highest spin vector $\O_{+}$
\begin{equation}
\label{eq:b01}
B^{*}_{M}(\bmu)\O_{+}=\xi^{M} p_{M}\O_{+}.
\end{equation}
Moreover the Bethe vectors are not orthogonal and their norms depend on the twist\break parameter.
\begin{lemma}
\label{lem:6}
Consider integers $M_{1},M_{2}\geq 0$, $M=\min\{ M_{1},M_{2} \}$ and complex numbers $\bl=\{\l_{1},\ldots,\l_{M_{1}}\}$ and $\bmu=\{\mu_{1},\ldots,\mu_{M_{2}}\}$ such that $\bl\cap\bmu=\emptyset$. Then the inner products between Bethe vectors is given by
\begin{align}
\label{eq:b05}
\langle\Psi_{M_{1}}(\bl)\lvert\Psi_{M_{2}}(\bmu)\rangle &= p_{M_{1}} p_{M_{2}}\,\xi^{M_{1} + M_{2}} + \sum_{k=1}^{M} p^{(k)}_{M_{1}-k}\,p^{(k)}_{M_{2}-k}\; \xi^{M_{1}+M_{2}-2k} \times\notag\\ &\quad\times\sum_{i_{1}<\cdots<i_{k}}^{M_{1}} \sum_{j_{1}<\cdots<j_{k}}^{M_{2}} \langle \Psi_{k}(\l_{i_{1}},\ldots,\l_{i_{k}})|\Psi_{k}(\mu_{j_{1}},\ldots,\mu_{j_{k}})\rangle_{0},
\end{align}
where $\langle\cdot|\cdot\rangle_{0}=\langle\cdot|\cdot\rangle|_{\xi=0}$ denotes the corresponding inner product of the $\sl_{2}$-invariant Gaudin model \cite{sklyanin2}.
\end{lemma}
\begin{proof}
The Bethe vectors are obtained from the action of the B-operators \eqref{eq:54} on $\O_{+}$
\begin{equation*}
\Psi_{M}(\bmu) = B_{M}(\bmu)\O_{+} = p_{M} \, \xi^{M} \O_{+} + \sum_{k=0}^{M-1} p_{k}^{(M-k)} \,\xi^{k} \sum_{j_{1<\cdots<j_{M-k}}}^{M} \Psi_{M-k}(\mu_{j_{1}},\ldots,\mu_{j_{M-k}})_{0}
\end{equation*}
where $\Psi_{M}(\bmu)_{0}= B_{M}(\bmu)|_{\xi=0}\O_{+}$ are the corresponding Bethe vectors of the $\sl_{2}$-invariant Gaudin model. Then the inner product of the Bethe vectors of the twisted model are
\begin{align*}
\langle\Psi_{M_{1}}(\bl)\lvert\Psi_{M_{2}}(\bmu)\rangle &= p_{M_{1}} p_{M_{2}}\,\xi^{M_{1} + M_{2}} + \sum_{k=0}^{M_{1}-1}\sum_{l=0}^{M_{2}-1} p_{k}^{(M_{1}-k)} p_{l}^{(M_{2}-l)}\, \xi^{k+l}\times\\ 
\times\sum_{i_{1}<\cdots<i_{M_{1}-k}}^{M_{1}}& \sum_{j_{1}<\cdots<j_{M_{2}-l}}^{M_{2}} \langle \Psi_{M_{1}-k}(\l_{i_{1}},\ldots,\l_{i_{M_{1}-k}})|\Psi_{M_{2}-l}(\mu_{j_{1}},\ldots,\mu_{j_{M_{2}-l}})\rangle_{0}\\
&= p_{M_{1}} p_{M_{2}}\,\xi^{M_{1} + M_{2}} + \sum_{n=1}^{M-1} p_{M_{1}-n}^{(n)} p_{M_{1}-n}^{(n)} \xi^{M_{1}+M_{2}-2n}\times\\
&\times \sum_{i_{1}<\cdots<i_{n}}^{M_{1}} \sum_{j_{1}<\cdots<j_{n}}^{M_{2}} \langle \Psi_{n}(\l_{i_{1}},\ldots,\l_{i_{n}})|\Psi_{n}(\mu_{j_{1}},\ldots,\mu_{j_{n}})\rangle_{0}.
\end{align*}
\end{proof}
\noindent The norms of the Bethe vector follow from the previous lemma.
\begin{corollary}
\label{offshellnorms}
The norms of Bethe vectors $\Psi_{M}(\bmu)$ are given by
\begin{equation}
\label{eq:b07}
\|\Psi_{M}(\bmu)\|^{2} = \sum_{k=0}^{M} (p^{(k)}_{M-k})^{2}\,\xi^{2(M-k)} \sum_{\substack{i_{1}<\cdots<i_{k}\\j_{1}<\cdots<j_{k}}}^{M} \langle \Psi_{k}(\mu_{i_{1}},\ldots,\mu_{i_{k}})|\Psi_{k}(\mu_{j_{1}},\ldots,\mu_{j_{k}})\rangle_{0}.
\end{equation}
\end{corollary}

\section{Conclusion}
The Gaudin model based on the $\sl_{2}$ classical r-matrix with jordanian twist is studied. This system can be obtained as the semiclassical limit of the XXX spin chain deformed by the jordanian twist. 
Alternatively, applying a certain similarity transformation on the XXZ Gaudin model and using the scaling limit, Kulish  \cite{Kulishjs} was able to postulate the spectrum and the Bethe states of the system. 

The result of Kulish \cite{Kulishjs} that the spectrum of the system coincides with the one of the $\sl_{2}$-invariant model is demonstrated by full implementation of  the algebraic Bethe Ansatz. In order to construct the Bethe vectors it is necessary to consider the creation operators which are not homogeneous polynomials of one of the generators of the corresponding loop algebra. However, it was convenient to consider a more general set of operators $B^{(k)}_{M}(\mu_{1},\ldots,\mu_{M})$, in order to simplify the calculation of the commutators between the creation operators and the generating function of the integrals of motion $t(\l)$. These operators are symmetric functions of their arguments and they satisfy certain recursive relations. Their commutation relations with the generators of the loop algebra are given and they are essential in the main step of the algebraic Bethe Ansatz. The creation operators are the particular case $B_{M}(\mu_{1},\ldots,\mu_{M})=B^{(0)}_{M}(\mu_{1},\ldots,\mu_{M})$. Thus the commutation relations between the creation operators and $t(\l)$ are easily calculated. The corresponding Bethe vectors are defined by the action of the creation operators on the highest spin vector $\O_{+}$ and the spectrum of the system is determined. However, the Bethe vectors in this case, are not eigenstates of the global Cartan generator $h_{gl}$. 

Some properties of the creation operators are fundamental in calculating the inner products and the norms of the Bethe states. It is necessary to consider the dual creation operators $B^{\ast}_{M}(\mu_{1},\ldots,\mu_{M})$ obtained by using the dual Gaudin model. The Bethe vectors $\Psi_{M}(\mu_{1},\ldots,\mu_{M})$ are not orthogonal for different $M$'s. Moreover, contrary to the $\sl_{2}$-invariant case, the generating function of integrals of motion is not hermitian.

The well known relation \cite{Babujian1994ba,frenkel} between the off-shell Bethe vectors of the Gaudin models related to simple Lie algebras and the solutions of Knizhnik-Zamolodchikov equation \cite{Knizhnikpe}
\begin{equation}
\label{eq:c01}
\k \frac{\partial}{\partial z_{a}} \psi(z_{1},\ldots,z_{N}) = H^{(a)}\psi(z_{1},\ldots,z_{N})
\end{equation}
where $H^{(a)}$ are the Gaudin hamiltoneans \eqref{Gaudinham}, also holds for the KZ equation related to the $\sl_{2}$ classical r-matrix with the jordanian twist. This relation is obtained by considering a Bethe vector $\Psi(\vec{\mu} |\vec{z})$, where the corresponding Bethe equations are not imposed, and the integral representation of solutions to the Knizhnik-Zamolodchikov equation,
\begin{equation}
\label{eq:c02}
\psi(z_{1},\ldots,z_{N})=\oint\cdots\oint \phi(\vec{\mu} |\vec{z})
\Psi(\vec{\mu} |\vec{z})d\vec{\mu}
\end{equation}
where $\phi(\vec{\mu} |\vec{z})$ is a scalar function
\begin{equation}
\label{eq:c03}
\phi(\vec{\mu} |\vec{z})=\prod_{i<j}^{M}(\mu_{i}-\mu_{j})^{\tfrac{4}{\k}} \prod_{a<b}^{N}(z_{a} - z_{b})^{\tfrac{\ell_{a}\ell_{b}}{\k}}
\prod_{a=1}^{N}\prod_{k=1}^{M}(z_{a} - \mu_{k})^{-\tfrac{2\ell_{a}}{\k}}.
\end{equation}
The partial derivatives of  the scalar factor \eqref{eq:c03} are simply written
\begin{equation}
\label{Phipartials}
\k\,\partial_{z_{a}}(\phi) = E_{M}^{(a)}\phi \quad\text{and}\quad \k\,\partial_{\mu_{i}}(\phi) = -2\rho_{M}(\mu_{i};\bmu^{(i)})\phi.
\end{equation}
It is a simple matter to check that $\psi$ given by \eqref{eq:c02} satisfies \eqref{eq:c01}. Due to the Leibniz rule $\partial_{z_{a}}(\phi\Psi) = \partial_{z_{a}}(\phi)\Psi + \phi\partial_{z_{a}}(\Psi)$ and 
the residue of \eqref{tcomPsi} at ~$\l = z_a$
\begin{equation}
\label{eq:c06}
H^{(a)}\Psi (\bmu) = E_{M}^{(a)}\Psi (\bmu) - 2 \sum_{i=1}^{M}\Psi_{M}^{(i,a)}(\bmu) \rho_{M}(\mu_{i};\bmu^{(i)})
\end{equation}
where $\Psi_{M}^{(i,a)} (\bmu) = \xm_{a}(\mu_{i})B^{(1)}_{M-1}(\bmu^{(i)})\O_{+}$ and $\xm_{a}(\mu_{i})=\frac{X_{a}^{-}}{ \mu_{i} - z_{a}} - \frac{\xi}{2} h_{a}$, we have 
\begin{equation}
\label{eq:c07}
\begin{split}
\k\,\partial_{z_a}(\phi \Psi) &= H^{(a)}(\phi\Psi) + 2\phi\sum_{i=1}^{M}\Psi_{M}^{(i,a)} (\bmu) \rho_{M}(\mu_{i};\bmu^{(i)}) + \k\,\phi\,\partial_{z_a}(\Psi)\\
&= H^{(a)}(\phi\Psi) - \k\sum_{i=1}^{M}\partial_{\mu_{i}}\left(\phi\Psi_{M}^{(i,a)}\right).
\end{split}
\end{equation}
To obtain the final formula we have used the lemma \ref{lem:2a} and formulas \eqref{Phipartials}.
Moreover, a closed contour integration of $\phi\Psi$ with respect to $\mu_{1},\ldots,\mu_{M}$ will cancel the contribution from the sum and therefore $\psi(z_{1},\ldots,z_{N})$ given by \eqref{eq:c02} will satisfy Knizhnik-Zamolodchikov equation.

\section*{Acknowledgments}
We wish to thank P. Kulish for enlightening discussions and helpful comments on the manuscript.
This work was supported by the projects POCTI/MAT/33858/2000,\break  POCTI/MAT/58452/2004 and the fellowship SFRH/BD/17771/2004.

\appendix
\section{Appendix: proofs of lemmas}

\textbf{Proof of lemma \ref{lem:1}}.
The recurrence relations in \textbf{ii.} and \textbf{iv.} are evident from the definition \ref{def:1}.\\ 
\textbf{i.} Given a fixed integer $k$, induction on $M$ is used. Consider $M=2$ in definition \eqref{BMk}:
\begin{equation}
\label{bola}
B^{(k)}_{2}(\mu_{1},\mu_{2}) = \xm(\mu_{1})\xm(\mu_{2}) + (k+1)\,\xi \xm(\mu_{1}) + k\,\xi \xm(\mu_{2}) + k(k+1)\xi^{2},
\end{equation}
then using \eqref{eq:05} it is straightforward to check that $B^{(k)}_{2}(\mu_{1},\mu_{2}) = B^{(k)}_{2}(\mu_{2},\mu_{1})$.

Assume $B^{(k)}_{N}(\mu_1,\ldots,\mu_{N})$ is symmetric for $M-1\geq N$. For $1\leq i<j<M$ it is clear from \textbf{ii.}  that
$$
B^{(k)}_{M}(\mu_1,\ldots,\mu_{i},\ldots,\mu_{j},\ldots,\mu_{M})=B^{(k)}_{M}(\mu_1,\ldots,\mu_{j},\ldots,\mu_{i},\ldots,\mu_{M}).
$$
The symmetry of $B^{(k)}_{M}(\mu_1,\ldots,\mu_{M})$ with respect to $\mu_{M-1}$ and $\mu_{M}$ must be shown. To this end the recurrence relation in \textbf{ii.} is to be iterated twice and the appropriate terms combined
\begin{align*}
& B^{(k)}_{M}(\mu_{1},\ldots,\mu_{M}) = B^{(k)}_{M-2}(\mu_{1},\ldots,\mu_{M-2})\left(\xm(\mu_{M-1}) + (M + k - 2)\xi \right)\\
& \x \left(\xm(\mu_{M}) + (M + k - 1)\xi \right)= B^{(k)}_{M-1}(\mu_{1},\ldots,\mu_{M-2},\mu_{M})\left(\xm(\mu_{M-1}) + (M+k-2)\xi \right)\\
& + B^{(k)}_{M-2}(\mu_{1},\ldots,\mu_{M-2})\left(\left[\xm(\mu_{M-1}),\xm(\mu_{M})\right]+\xi\left(\xm(\mu_{M-1}) + (M+k-2)\xi \right)\right)\\
&= B^{(k)}_{M-1}(\mu_{1},\ldots,\mu_{M-2},\mu_{M})\left(\xm(\mu_{M-1}) + (M+k-2)\xi \right)
+ \xi B^{(k)}_{M-1}(\mu_{1},\ldots,\mu_{M-2},\mu_{M})\\
&=B^{(k)}_{M}(\mu_{1},\ldots,\mu_{M-2},\mu_{M},\mu_{M-1}).
\end{align*}
\textbf{iii.} Applying the induction on $M$, for a fixed integer $k$. Set $M=2$, it is a direct  consequence of \eqref{bola} that
$$
B^{(k)}_{2}(\mu_{1},\mu_{2}) = B^{(k-1)}_{2}(\mu_{1},\mu_{2}) +\xi (B^{(k)}_{1}(\mu_{1}) + B^{(k)}_{1}(\mu_{2})).
$$
Assume \textbf{iii.} is true for $M-1$ then 
\begin{align*}
&B^{(k)}_{M}(\bmu) = \left(B^{(k-1)}_{M-1}(\bmu^{(M)}) + \xi\sum_{i=1}^{M-1}B^{(k)}_{M-2}(\bmu^{(i,M)})\right)(\xm(\mu_{M}) + (M+k-1)\xi) \\
&=B^{(k-1)}_{M}(\bmu) + \xi\, B^{(k-1)}_{M-1}(\bmu^{(M)}) + \xi\,\sum_{i=1}^{M-1}\left(B^{(k)}_{M-1}(\bmu^{(i)}) + \xi\, B^{(k)}_{M-2}(\bmu^{(i,M)})\right)\\
&=B^{(k-1)}_{M}(\bmu) + \xi\,\sum_{i=1}^{M-1}B^{(k)}_{M-1}(\bmu^{(i)}) + \xi B^{(k)}_{M-1}(\bmu^{(M)}) = B^{(k-1)}_{M}(\bmu) + \xi\sum_{i=1}^{M}B^{(k)}_{M-1}(\bmu^{(i)}).
\end{align*}
\qed

\noindent
\textbf{Proof of lemma \ref{lem:2}}.
From definition \ref{def:1} $B^{(k)}_{1}(\mu) = \xm(\mu)+k\xi$, and commutators \eqref{eq:05} it is clear that $M=1$ in (2.26-28) is given by
$$
\left[h(\l),B^{(k)}_{1}(\mu)\right] = \left[h(\l),\xm(\mu)\right] = 2\frac{B^{(k)}_{1}(\l) - B^{(k)}_{1}(\mu)}{\l - \mu} + \xi \betheop_{1}(\mu),
$$
$$
\left[\xp(\l),B^{(k)}_{1}(\mu)\right] = \left[\xp(\l),\xm(\mu)\right] = -\frac{\betheop_{1}(\l) - \betheop_{1}(\mu)}{\l - \mu} + \xi \xp(\l),
$$
$$
\left[\xm(\l),B^{(k)}_{1}(\mu)\right] = \left[\xm(\l),\xm(\mu)\right] = \xi B^{(k)}_{1}(\mu) - \xi B^{(k)}_{1}(\l).
$$
The induction method is used to demonstrate lemma \ref{lem:2}. Assume that (2.26-28) hold for $B_{N}^{(k)}(\bmu),\, M-1\geq N\geq 1$ then, to show that these formulas are valid for $M$, the recurrence relation \textbf{ii.} in lemma \ref{lem:1} is used 
\begin{align*}
&\left[h(\l),B^{(k)}_{M}(\bmu)\right] = \left[h(\l),B^{(k)}_{M-1}(\bmu^{(M)})\right](B_{1}(\mu_{M})+(M+k-1)\xi) + B^{(k)}_{M-1}(\bmu^{(M)})\\
& \x [h(\l),\xm(\mu_{M})] = \sum_{i=1}^{M-1}\left(2\frac{B^{(k)}_{M}(\{\l\}\cup\bmu^{(i)})-B^{(k)}_{M}(\bmu)}{\l-\mu_{i}} +\xi B^{(k+1)}_{M-1}(\bmu^{(i)})\betheop_{M-1}(\mu_{i};\bmu^{(i,M)})\right)\\
& + \xi \sum_{i=1}^{M-1} B^{(k+1)}_{M-2}(\bmu^{(i,M)})[h(\mu_{i}),\xm(\mu_{M})] + B^{(k)}_{M-1}(\bmu^{(M)})\left(2\frac{B_{1}(\l) - B_{1}(\mu_{M})}{\l - \mu_{M}} + \xi h(\mu_{M})\right)\\
& = 2\sum_{i=1}^{M}\frac{B^{(k)}_{M}(\{\l\}\cup\bmu^{(i)})-B^{(k)}_{M}(\bmu)}{\l-\mu_{i}} +\xi \sum_{i=1}^{M-1} B^{(k+1)}_{M-1}(\bmu^{(i)})\left(\betheop_{M-1}(\mu_{i};\bmu^{(i,M)})+\frac{2}{\mu_{M}-\mu_{i}}\right)\\
&+ \xi B^{(k+1)}_{M-1}(\bmu^{(M)})\sum_{i=1}^{M-1} \frac{2}{\mu_{i}-\mu_{M}} +\xi\left(B^{(k)}_{M-1}(\bmu^{(M)})+\xi\sum_{i=1}^{M-1}B^{(k+1)}_{M-2}(\bmu^{(i,M)})\right)h(\mu_{M})\\
&= \sum_{i=1}^{M}\left(2\frac{B^{(k)}_{M}(\l\cup\bmu^{(i)})-B^{(k)}_{M}(\bmu)}{\l-\mu_{i}} + \xi B^{(k+1)}_{M-1}(\bmu^{(i)})\betheop_{M}(\mu_{i};\bmu^{(i)})\right),
\end{align*}
here where used, where appropriate, the induction hypothesis, the commutators $[h(\l),\xm(\mu)]$ and properties \textbf{ii.} and \textbf{iv.} in lemma \ref{lem:1}. Thus \eqref{eq:27} is proved.
\begin{align*}
&\left[\xp(\l),B^{(k)}_{M}(\bmu)\right] =\left[\xp(\l),B^{(k)}_{M-1}(\bmu^{(M)})\right](B_{1}(\mu_{M})+(M+k-1)\xi) + B^{(k)}_{M-1}(\bmu^{(M)})\\
&\x \left[\xp(\l),\xm(\mu_{M})\right] = - \sum_{i=1}^{M-1} B^{(k+1)}_{M-1}(\bmu^{(i)}) \frac{\betheop_{M-1}(\l;\bmu^{(i,M)})-\betheop_{M-1}(\mu_{i};\bmu^{(i,M)})}{\l-\mu_{i}}\\
& - 2\sum_{i<j}^{M-1}\frac{B^{(k+1)}_{M-1}(\{\l\}\cup\bmu^{(i,j)})}{(\l-\mu_{i})(\l-\mu_{j})} + \xi \sum_{i=1}^{M-1} B^{(k+1)}_{M-1}(\bmu^{(i)})\xp(\l)\\
&+\left( B^{(k)}_{M-1}(\bmu^{(M)})+\xi\sum_{i=1}^{M-1}B^{(k+1)}_{M-2}(\bmu^{(i,M)}) \right) \left(-\frac{h(\l)-h(\mu_{M})}{\l-\mu_{M}} + \xi\xp(\l)\right)\\
& - \sum_{i=1}^{M-1} B^{(k+1)}_{M-2}(\bmu^{(i,M)}) \frac{\left[h(\l),\xm(\mu_{M})\right]-\left[h(\mu_{i}),\xm(\mu_{M})\right]}{\l-\mu_{i}}\\
&= - \sum_{i=1}^{M-1} B^{(k+1)}_{M-1}(\bmu^{(i)}) \frac{\betheop_{M}(\l;\bmu^{(i)})-\betheop_{M}(\mu_{i};\bmu^{(i)})}{\l-\mu_{i}} - 2\sum_{i<j}^{M-1}\frac{B^{(k+1)}_{M-1}(\{\l\}\cup\bmu^{(i,j)})}{(\l-\mu_{i})(\l-\mu_{j})}\\
&+\xi \sum_{i=1}^{M} B^{(k+1)}_{M-1}(\bmu^{(i)})\xp(\l) - B^{(k+1)}_{M-1}(\bmu^{(M)})\frac{\betheop_{M}(\l;\bmu^{(M)})-\betheop_{M}(\mu_{M};\bmu^{(M)})}{\l-\mu_{M}}\\
& - 2\sum_{i=1}^{M-1} \frac{B^{(k+1)}_{M-1}(\{\l\}\cup\bmu^{(i,M)})}{(\l-\mu_{i})(\l-\mu_{M})}\\
&=- \sum_{i=1}^{M} B^{(k+1)}_{M-1}(\bmu^{(i)}) \frac{\betheop_{M}(\l;\bmu^{(i)})-\betheop_{M}(\mu_{i};\bmu^{(i)})}{\l-\mu_{i}} - 2\sum_{i<j}^{M}\frac{B^{(k+1)}_{M-1}(\{\l\}\cup\bmu^{(i,j)})}{(\l-\mu_{i})(\l-\mu_{j})}\\
&\quad + \xi \sum_{i=1}^{M} B^{(k+1)}_{M-1}(\bmu^{(i)})\xp(\l),
\end{align*}
here we used, where appropriate, the commutators $[h(\l),\xm(\mu)]$ and $[\xp(\l),\xm(\mu)]$. Thus \eqref{eq:28} is proved.
\begin{align*}
&\left[\xm(\l),B^{(k)}_{M}(\bmu)\right] = \left[\xm(\l),B^{(k)}_{M-1}(\bmu^{(M)})\right](B_{1}(\mu_{M})+(M+k-1)\xi)+ B^{(k)}_{M-1}(\bmu^{(M)})\\
&\x\left[\xm(\l),B_{1}(\mu_{M})\right]\ = \Biggl((M-1)\,\xi \Biggr.\left. B^{(k)}_{M-1}(\bmu^{(M)}) - \xi\sum_{i=1}^{M-1} B^{(k)}_{M-1}(\{\l\}\cup\bmu^{(i,M)})\right)\\
& \x (B_{1}(\mu_{M})+(M+k-1)\xi)+ \xi B^{(k)}_{M-1}(\bmu^{(M)})(B_{1}(\mu_{M})-B_{1}(\l))\\
& = M\,\xi\, B^{(k)}_{M}(\bmu) - \xi\sum_{i=1}^{M} B^{(k)}_{M}(\{\l\}\cup\bmu^{(i)}),
\end{align*}
here we used, where appropriate, the commutators $[\xm(\l),\xm(\mu)]$. \qed

\noindent
\textbf{Proof of lemma \ref{lem:2a}}.
In the case for $M=1$ the proof is straightforward
$$
\frac{\partial}{\partial z_{a}} B_{1}(\mu)= \frac{\xm_{a}}{(\mu - z_{a})^{2}} = - \frac{\partial}{\partial \mu}\left(\frac{\xm_{a}}{\mu_{i} - z_{a}}-\frac{\xi}{2}h_{a}\right)= - \frac{\partial}{\partial \mu}\xm_{a}(\mu_{i}).
$$
Assume \eqref{eq:2a} is true for any set $\bmu$ with $|\bmu |<M$ complex numbers. Use formula  \eqref{eq:22} and the induction hypothesis together with 1. of lemma \ref{lem:1} we can write 
\begin{align*}
\frac{\partial}{\partial z_{a}} B_{M}(\bmu) &= \frac{\partial}{\partial z_{a}} \left(B_{M-1}(\bmu^{(M)})\left(B_{1}(\mu_{M}) + (M-1)\xi \right)\right)\\
&=-\sum_{i=1}^{M-1}\frac{\partial}{\partial \mu_{i}}\left(\xm_{a}(\mu_{i})B^{(1)}_{M-1}(\bmu^{(i)})\right)- \frac{\partial}{\partial \mu_{M}}B_{M-1}(\bmu^{(M)})\xm_{a}(\mu_{M})\\
&=-\sum_{i=1}^{M-1}\frac{\partial}{\partial \mu_{i}}\left(\xm_{a}(\mu_{i})B^{(1)}_{M-1}(\bmu^{(i)})\right)\\
&\quad - \frac{\partial}{\partial \mu_{M}}\left(\xm_{a}(\mu_{M})\, B_{M-1}(\bmu^{(M)})+ \frac{\left[B_{M-1}(\bmu^{(M)}),\xm_{a}\right]}{\mu_{M} - z_{a}}\right)\\
&=-\sum_{i=1}^{M}\frac{\partial}{\partial \mu_{i}}\left(\xm_{a}(\mu_{i})B^{(1)}_{M-1}(\bmu^{(i)})\right).
\end{align*}
\qed

\newpage

\end{document}